\documentclass[numbers,square,sort&compress,review]{elsarticle}
\journal{Nucl. Instrum. Methods Phys. Res. A}
\graphicspath{{figures/}}
\usepackage{amsmath,amssymb}
\usepackage[colorlinks=true]{hyperref}

\begin{document}
%=====================================================================

%--------------------------------------------------------------------- 
\begin{frontmatter}
%--------------------------------------------------------------------- 
  %% \textcolor{red}{Revision Draft of \today}\\
  \title{Novel optical interferometry of synchrotron radiation for absolute
    electron beam energy measurements}

  \author[kph]{P.~Klag\fnref{PhD}}
  \author[kph]{P.~Achenbach\corref{corr}}\ead{achenbach@uni-mainz.de}
  \author[kph]{M.~Biroth}
  \author[tohoku]{T.~Gogami\fnref{GPPU}}
  \author[kph]{P.~Herrmann}
  \author[tohoku]{M.~Kaneta}
  \author[tohoku]{Y.~Konishi}
  \author[kph]{W.~Lauth}
  \author[tohoku]{S.~Nagao}
  \author[tohoku]{S.~N.~Nakamura}
  \author[kph]{J.~Pochodzalla}
  \author[kph]{J.~{Roser}}
  \author[tohoku]{Y.~Toyama\fnref{GPPU}}
  \fntext[PhD]{Part of doctoral thesis.}
  \fntext[GPPU]{Graduate Program on Physics for the Universe, Tohoku
    University (GP-PU)} 
	
  \address[kph]{Institut f\"ur Kernphysik, Johannes
    Gutenberg-Universit\"at, 55099 Mainz, Germany}
  \address[tohoku]{Department of Physics, Graduate School of Science,
    Tohoku University, Sendai, 980-8578, Japan}

  \cortext[corr]{Corresponding author.}

\begin{abstract}
  A novel interferometric method is presented for the measurement of
  the absolute energy of electron beams. In the year 2016, a
  pioneering experiment was performed using a 195\,MeV beam of the
  Mainz Microtron (MAMI). The experimental setup consisted of two
  collinear magnetic undulators as sources of coherent optical
  synchrotron light and a high-resolving grating monochromator. Beam
  energy measurements required the variation of the relative undulator
  distance in the decimeter range and the analysis of the intensity
  oscillation length in the interference spectrum.  A statistical
  precision of 1\,keV was achieved in 1 hour of data taking, while
  systematic uncertainties of 700\,keV were present in the
  experiment. These developments aim for a relative precision of
  $10^{-5}$ in the absolute momentum calibrations of spectrometers and
  high-precision hypernuclear experiments. Other electron accelerators
  with beam energies in this regime such as the Mainz Energy
  Recovering Superconducting Accelerator (MESA) might benefit from
  this new method.
\end{abstract}


%--------------------------------------------------------------------- 
\end{frontmatter}
%--------------------------------------------------------------------- 

%\linenumbers 

%--------------------------------------------------------------------- 
\section{Introduction}
%--------------------------------------------------------------------- 

During the last years, a new method of decay-pion spectroscopy was
pioneered at the Mainz Microtron (MAMI), which has the potential to
achieve mass measurements of several light hypernuclei with a
precision better than 50\,keV$\!/c^2$~\cite{Esser:2015,Schulz:2016}.
Such a high precision is indeed required, {\em e.g.}, for the
determination of the spin dependence of the charge symmetry breaking
effect in light hypernuclei~\cite{Achenbach:2016}. Furthermore, a
planned precision measurement of the mass of lightest hypernucleus,
composed of a proton, a neutron, and a $\Lambda$-particle, will address
the so-called hypertriton puzzle~\cite{HiHyp:2018}. Presently, the
largest systematic error in these experiments originated from the
uncertainty in the MAMI beam energy affecting the absolute momentum
calibration of the spectrometers by $\delta p \approx \pm$
100\,keV$\!/c$, the sum of all other systematic errors contributed one
order of magnitude less~\cite{Schulz:2016}.
 
In this work, a novel interferometric method is presented for the
measurement of the absolute energy of electron beams in the range of
100 to 200\,MeV. The method is based on the analysis of the intensity
oscillation length in the synchrotron spectrum from two collinear
sources, thus reducing the energy determination to a relative distance
measurement in the decimeter range and the spectroscopy of a narrow
optical wavelength band.

The paper is organized as follows. After introducing different methods
for the energy determinations at electron accelerators in
Section~\ref{sec:accelerators}, the MAMI accelerator is briefly reviewed
in Section~\ref{sec:MAMI} with a focus on its energy stability and
absolute energy determination. The operating principle of the novel
method is presented in Section~\ref{sec:interferometry}. In
Section~\ref{sec:setup}, the experimental setup used for the
pioneering experiment at MAMI is described. Images of the synchrotron
radiation from different measurements are shown in
Section~\ref{sec:images}. Results from the evaluations of the spectra
and the determination of the MAMI beam energy are shown in
Section~\ref{sec:beamenergy}. A conclusion follows in
Section~\ref{sec:conclusion}.

%--------------------------------------------------------------------- 
\section{Energy determinations at electron accelerators}
\label{sec:accelerators}
%--------------------------------------------------------------------- 

In storage rings, the beam energy can be measured with a relative
uncertainty of few $10^{-3}$ from the integrated dipole field along
the ring~\cite{Sun:2009}. At some facilities, {\em e.g.}, at the
VEPP-4M collider at BINP and at the SPEAR3 electron storage ring, more
precise determinations have been achieved with the resonant spin
depolarization technique.  Relative uncertainties for the energy
measurement on the order of $10^{-5}$ were realized for
VEPP-4M~\cite{Blinov:2009} and on the order of $3 \times 10^{-6}$ for
SPEAR3~\cite{Wootton:2012}.  The application of this method is limited
to spin-polarized beams in high-energy storage rings and therefore
cannot be used at MAMI, in which the beam is passing the accelerator
only once.

The Compton backscattering method does not require a polarized beam
and can be used in a wide range of beam energies from a few hundred
MeV to a few GeV. The relative uncertainty of this method is usually
on the order of $10^{-4}$~\cite{Sun:2009,Klein:1997,Klein:2002}. In
these measurements, beam particles are collided head-on with photons
from a laser. The maximum energy $E_\gamma^{\text{max}}$ of the
backscattered Compton $\gamma$-rays is measured with high-purity
germanium detectors and converted into the central primary beam
energy. The systematic uncertainty of the method is dominated by the
absolute calibration of the energy scale of the detector for the
$\gamma$-ray. The Compton backscattering of laser photons realized at
BESSY I and BESSY II has reached accuracies of $\delta E/E = 5 \times
10^{-5}$ at 1\,718\,MeV~\cite{Klein:2002} and $2 \times 10^{-4}$ at a
lower energy of 800\,MeV~\cite{Klein:1997}. With the same method, a
relative systematic uncertainty of $\delta E/E = 2 \times 10^{-5}$ was
achieved for the 1\,840\,MeV beam at BEPC-II~\cite{Zhang:2012}.

The application of the method to lower beam energies is challenging
because of the continuous decrease of the Compton edge with decreasing
beam energy:
\begin{align}
  E_\gamma^{\text{max}} = \frac{4\gamma^2 E_\lambda}{1+4\gamma^2
    E_\lambda/E_{\text{beam}}} \approx 4\gamma^2 E_\lambda\,,
\end{align}
where $E_\lambda$ is the energy of the laser photon and $\gamma$ the
Lorentz factor of the beam. When colliding laser photons of 800\,nm
with an electron beam of 500\,MeV, the resulting energy spectrum
extends to $E_\gamma^{\text{max}} \sim 6$\,MeV which can be determined
with the best possible calorimeters with an uncertainty of a few keV,
resulting in a theoretical resolution of a few $10^{-4}$. For a beam
energy of 195\,MeV this theoretical best resolution increases to above
few $10^{-3}$.  Furthermore, the $\gamma$-ray collimation as well as
the finite electron beam emittance impacts on the $\gamma$-ray
spectrum. Under certain beam conditions, the determination of the beam
energy from the spectrum is significantly influenced~\cite{Sun:2009}.

To overcome these limitations, a new method is developed for the low
energy electron beams at MAMI. Other electron accelerators with beam
energies in this regime such as the Mainz Energy Recovering
Superconducting Accelerator (MESA), currently under construction,
might benefit from this work. MESA will consist of two cryo-modules
with an acceleration capacity of 25\,MeV each and three recirculation
arcs for a maximum beam energy of 155\,MeV. The MESA beam energy will
be stabilized using the return arc with maximum longitudinal
dispersion and two beam phase cavity monitors.  Because of high
demands from the experiments, among them the detection of order
$10^{-8}$ parity-violating cross section asymmetries in electron
scattering, the beam energy fluctuations need to be minimized to
unprecedented low levels and the absolute beam energy needs to be
determined with high precision.

%--------------------------------------------------------------------- 
\section{The MAMI electron accelerator}
\label{sec:MAMI}
%--------------------------------------------------------------------- 

MAMI is a multi-stage accelerator based on normal conducting
radio-frequency (rf) cavities that can deliver a continuous-wave (cw)
electron beam~\cite{Herminghaus:1976,Kaiser:2008,Dehn:2011}. Electrons
are drawn from the source with a static high voltage of 100\,kV and
are further accelerated by an injector linear accelerator (linac) to
an energy of 3.5\,MeV, reaching relativistic velocities of $\beta >$
0.99. The recirculating part consists of three cascaded racetrack
microtrons (RTMs) and an additional harmonic double-sided microtron
(HDSM) as a fourth stage. In each RTM, the beam is recirculated
through two homogeneous 180$^\circ$ dipole magnets to a common linac
section composed of a series of axially coupled accelerating cavities
that are powered by several klystrons using a rf of 2.45\,GHz. The
first two RTMs accelerate the beam to 14.9\,MeV and 180\,MeV,
respectively. The third RTM has 90 return paths to the linac section
and the beam can be extracted from all even-numbered paths, so that
this stage has a final energy from 180 to 855\,MeV in 15\,MeV
steps. The beam intensity is limited by the available rf power to a
maximum current of 100\,$\mu$A.  The HDSM consists of two normal
conducting linacs through which the electrons are guided up to 43
times by a pair of 90$^\circ$-bending magnets at each end. For stable
beam dynamics, the linacs operate at the harmonic frequencies of 4.90
and 2.45\,GHz.  This stage can deliver a beam with energies of up to
1.6\,GeV.

The energy spread of a typical beam from RTM3 is dominated by the
stochastic emission of synchrotron radiation photons.  This energy
loss per turn grows with the third power of the beam energy.
Fortunately, the strong longitudinal focusing in RTMs compensates
synchrotron radiation losses in each turn by a proper phase migration.
Residual rf phase and amplitude fluctuations have only little
influence on the beam energy.  The remaining width is
$\sigma_{\text{beam}} <$ 13\,keV corresponding to a relative energy
spread of $\Delta \gamma/\gamma = 1.5 \times 10^{-5}$ when expressing
the beam energy by its Lorentz factor $\gamma = E_{\text{beam}} / m_e
c^2$.
 
An excellent energy stability with a very small drift over time of
less than 1\,keV has been realized by a combination of two digital
feedback loops~\cite{Seidl:2000}. A fast loop eliminates output energy
deviations by acting on the rf phase using the time-of-flight
dependence of bunches from the last return path to the extraction beam
line. A slow loop stabilizes the measured tune of the RTM3 by small
changes of the linac amplitude. ~\cite{Euteneuer:1994}

The absolute beam energy can be measured using magnetic spectrometry
inside the RTM3 stage of the accelerator by exact determination of the
beam position on the linac axis and in a higher (73rd) return
path. The main instrumentation is a 9.8\,GHz XY beam position monitor
(XYMO), whose transverse separation of its electrical center to the
linac axis is known with a precision of approximately 0.4\,mm. The
resolution of the monitor is much higher than 0.1\,mm in diagnostic
pulse mode.  For the energy measurements, the beam is first centered
on the linac axis and then centered with the use of calibrated
correction steerer magnets on the XYMO axis. From the correction
currents, the bending radius of the beam in the 73rd turn can be
calculated.  The magnetic field $B$ inside the RTM3 dipoles is known
by NMR measurements and the field accuracy $\delta B/B$ is on the
order of $10^{-4}$. The total uncertainty of the beam energy $\delta
E_{73}$ at $E_{73} \approx 727$\,MeV is 120\,keV including
contributions from geodetic measurement errors, calibration errors of
the steerers, and dominant angle errors.  By use of the well
established and benchmarked particle tracking program PTRACE, the beam
energy $E_n$ of the extracted turn number $n$ can be interpolated from
the value $E_{73}$. The uncertainty for $E_n$ is 130\,keV when
including a systematic error from the interpolation on the order of
55\,keV. A conservative error estimation for the absolute energy of
MAMI including rf phase and amplitude errors leads to a total accuracy
of $\delta E_{\text{beam}} = 160$\,keV~\cite{MAMI-03-06:2006}.

%--------------------------------------------------------------------- 
\section{Interferometry of synchrotron radiation}
\label{sec:interferometry}
%--------------------------------------------------------------------- 

\begin{figure}[tbp]
  \centering
  \includegraphics[width=\textwidth]{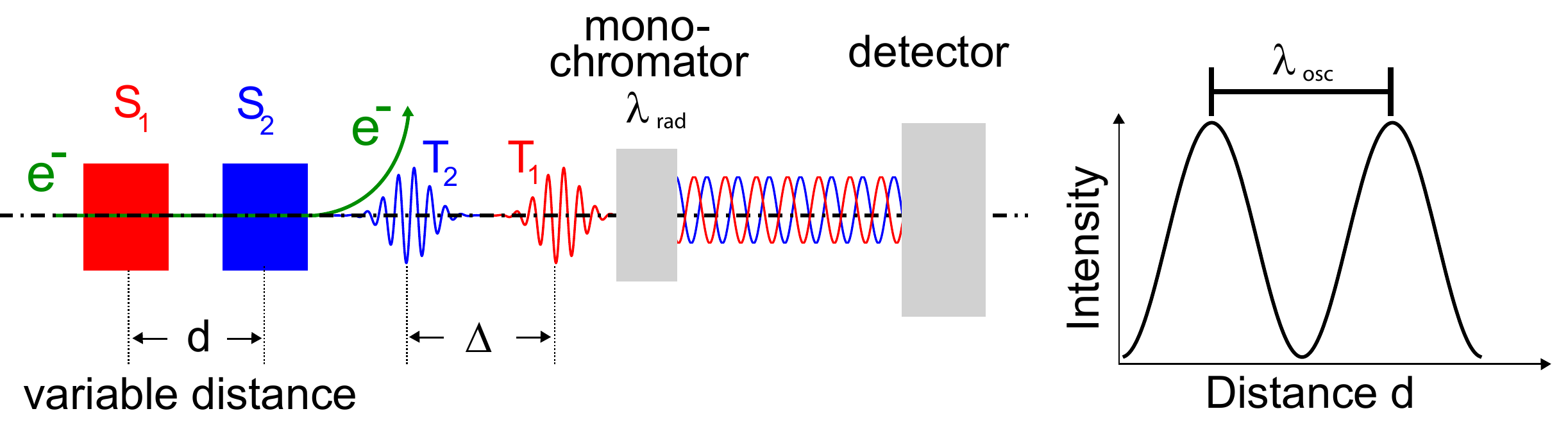}
  \caption{Schematic drawing (not to scale) of the novel method for
    absolute beam energy measurements comprising two spatially
    separated sources of coherent light at an electron beam and an
    optical interferometer system.  Relativistic electrons ($e^-$)
    pass through the two sources ($S_1$ and $S_2$ separated along the
    axis by a variable distance $d$) and produce wave trains of
    coherent light ($T_1$ and $T_2$ separated by a difference
    $\Delta$). A monochromator serves as a Fourier analyzer of the
    wave trains and a position sensitive optical detector is used to
    observe the interference. The intensity for a selected wavelength
    $\lambda_{\text{rad}}$ varies periodically with the distance $d$
    as shown to the right.  The oscillation length
    $\lambda_{\text{osc}}$ is directly related to the Lorentz factor
    $\gamma = \sqrt{ \lambda_{\text{osc}} / 2\lambda_{\text{rad}}}$
    when observed on-axis.}
  \label{fig:expSetup}
\end{figure}

The method is based on interferometry with two spatial separated light
sources driven by relativistic
electrons~\cite{Dambach:1998,Dambach:1999,Lauth:2006}.  The basic idea
will be explained by means of the schematic drawing shown in
Fig.~\ref{fig:expSetup}. An electron beam with Lorentz factor $\gamma$
passes a pair of undulators S$_1$ and S$_2$ separated by a distance
$d$. Further details of the undulator pair can be found in
Section~\ref{subsec:undulators}. The succession of the wave trains
T$_1$ and T$_2$ at the exit of the undulator pair is opposite to the
order of the two sources because the electron velocity $v$ is slower
than the speed of light.  These trains are separated along the axis by
the distance
\begin{align}
  \Delta(\theta,d) = \left( \frac{2+K^2}{4\gamma^2} +
    \frac{\theta^2}{2} \right)L_U + \left(
    \frac{1}{2\gamma^2}+\frac{\theta^2}{2} \right)d\,,
\end{align}
which is a linear function in $d$. Details on the superposition of the
two wave trains are given in~\cite{Dambach:1998}.  The slope is only
dependent on the Lorentz factor $\gamma$ and the observation angle
$\theta$ with respect to the electron beam direction. The
dimensionless undulator parameter is $K = (e/2\pi m_e c) \cdot B_0
\cdot \lambda_U$ and $L_U \simeq n \lambda_U$ is the length of the
undulator with $\lambda_U$ the undulator period and $n$ is the number
of periods. The undulators act as sources for the emission of coherent
light with the amplitudes $A_{1,2}$ of the two wave trains having a
phase difference of $\phi(\theta,d) = 2\pi \Delta(\theta,d) /
\lambda_{\text{rad}}$ for a selected wavelength of the radiation. The
intensity $I = A^2$ of the two interfering sources is given by:
\begin{align}
  I(\theta,d) = |A_1|^2 + |A_2|^2 + 2|A_1||A_2| \cos {\frac{2\pi
      \Delta(\theta,d)}{\lambda_{\text{rad}}}}\,.
\end{align}
A monochromator can serve as a Fourier analyzer of the wave trains. If
both wave trains interfere in a position sensitive detector and $d$ is
varied by moving one of the sources, then the revolving phase
$\phi(\theta,d)$ can be observed as intensity oscillations with
oscillation length $\lambda_{\text{osc}} =
2\gamma^2\lambda_{\text{rad}}(1+\gamma^2\theta^2)^{-1}$. For a
wavelength $\lambda_{\text{rad}}$ selected by the monochromator and
on-axis observation at $\theta = 0$, the oscillation length
directly relates to the Lorentz factor $\gamma$,
\begin{align}
  \gamma^2 =
  \frac{1}{2}\frac{\lambda_{\text{osc}}}{\lambda_{\text{rad}}}\,,
\end{align}
which was first pointed out in~\cite{Dambach:1999}. Both
$\lambda_{\text{osc}}$ as well as $\lambda_{\text{rad}}$ can be
measured with very high precision.  The method is independent of the
nature of the emission process, provided that the produced light is
coherent.

%--------------------------------------------------------------------- 
\section{Experimental setup}
\label{sec:setup}
%--------------------------------------------------------------------- 

%--------------------------------------------------------------------- 
\subsection{Undulator pair}
\label{subsec:undulators}
%--------------------------------------------------------------------- 

\begin{figure}[tbp]
  \centering
  \includegraphics[width=\textwidth]{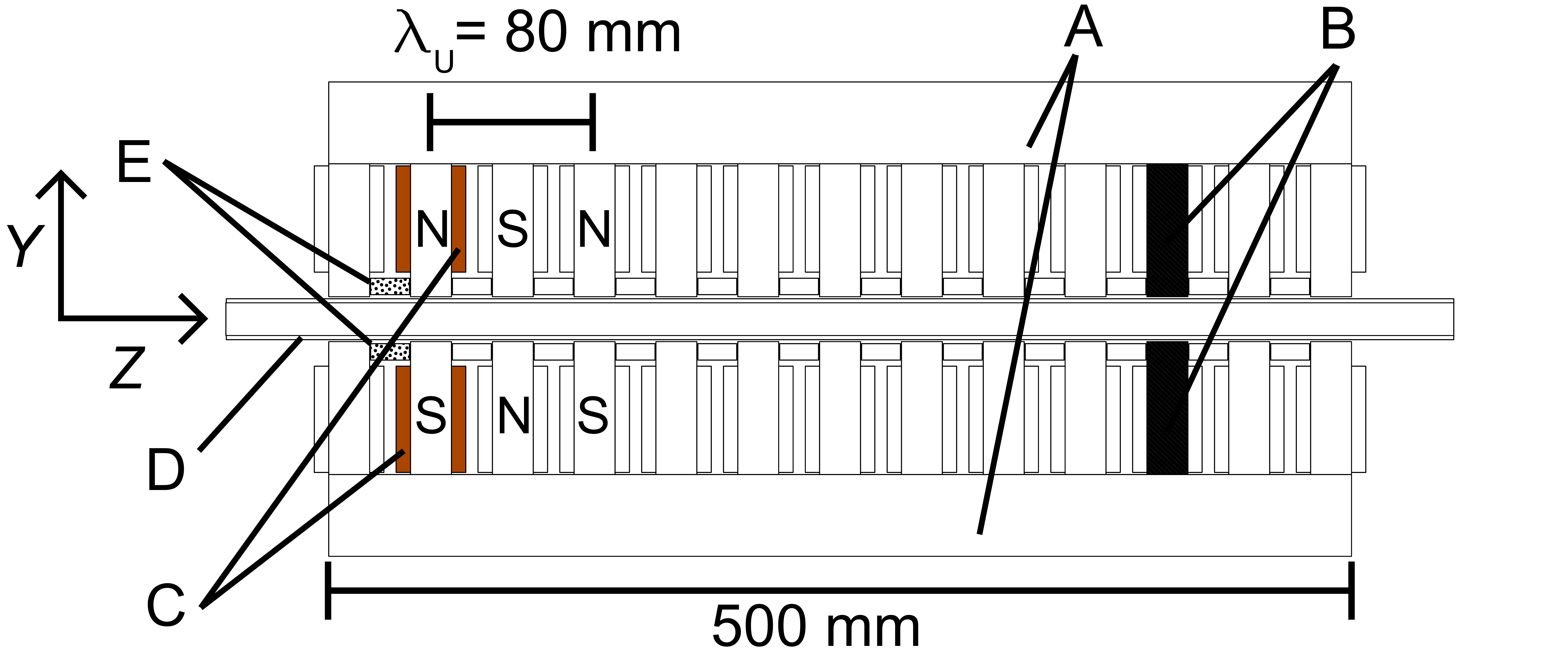}
  \caption{Schematic drawing of the cross section of the undulators in
    the middle plane. The yoke (A) of 500\,mm length connected the
    pole pieces, all parts made from Armco iron. Each of the 13
    alternating magnetic pole pairs (B) were driven by individually
    controlled currents through copper coils (C) that were separated
    from the beam pipe (D) by spacers (E).  The undulator period given
    by this geometry was $\lambda_U = 80$\,mm. A coordinate system was
    used in which $z$ is in beam direction, $y$ is in anti-gravity
    direction, and $x$ in transverse direction.}
  \label{fig:SchematicUndulator}
\end{figure}

\begin{figure}[tbp]
  \centering
  \includegraphics[width=\textwidth]{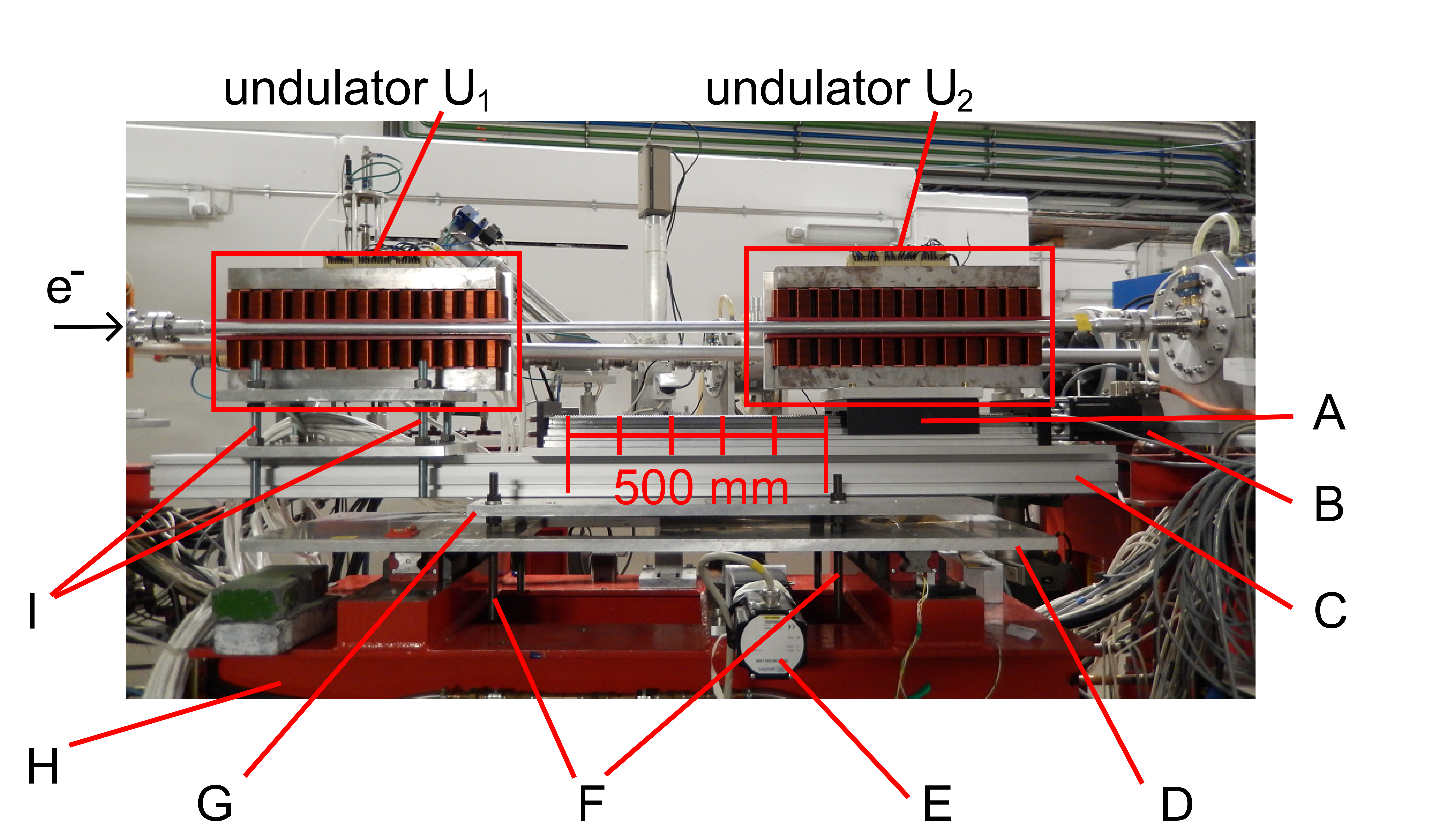}
  \caption{Photograph of the setup used for the pioneering experiment
    at the electron beam line, taken when the two undulators were
    separated by the maximum distance of $d = 500$\,mm.  The electron
    beam entered from the left. A high-precision linear stage (A)
    could move undulator U$_2$ away from undulator U$_1$ as far as
    500\,mm along the beam axis with a step motor (B). Longitudinal
    alignment of the undulator pair with respect to beam line elements
    was done by an aluminium profile (C) of 80\,mm height. An
    aluminium base (D) could be moved by a motor (E) perpendicular to
    the beam direction. The adjustment of the angle to the beam line
    was achieved with sub-mrad precision by metric fine threads (F)
    holding another aluminium base (G). A welded steel construction
    (H) supported the setup. The height of undulator~U$_1$ was aligned
    with respect to U$_2$ using four threads (I).}
  \label{fig:PhotoUndulators}
\end{figure}

To realize beam energy measurements in the range of 100 to 200\,MeV
with synchrotron radiation interferometry, an undulator pair with an
undulator parameter $K = 0.934 \cdot B_0 \text{[T]} \cdot \lambda_U
\text{[cm]} \sim 1$ and wavelengths in the visible range are
practical. Such wavelengths allow the use of high-resolving optical
monochromators and spectrometers. Fig.~\ref{fig:SchematicUndulator}
shows a drawing of the cross section of the undulators in
the middle plane. The number of periods in each undulator was $n =
13$. An undulator period of $\lambda_U = 80$\,mm and magnetic field
amplitudes of $B_0 = 130$\,mT were achieved with normal conducting
coils, where the current through the coil pairs could be individually
controlled.  The resulting undulator parameter was $K \approx 0.97$,
for which wavelengths near 400\,nm were expected in the synchrotron
spectrum.

Fig.~\ref{fig:PhotoUndulators} shows a photograph of the setup at the
electron beam line. A high-precision linear stage could move undulator
U$_2$ away from undulator U$_1$ a distance $d$ between 0 and 500\,mm
along the beam axis, thus covering approximately $4 \times
\lambda_{\text{osc}}$ in one beam energy measurement. The position was
given by a rotational sensor counting the turns of a spindle.  The
heights of the undulators were aligned with respect to each other. In
order to correct the orientation of the undulators against each other,
one of the undulators could be tilted.  The adjustment of the angle to
the beam line was achieved with sub-mrad precision by tilting the
aluminium base. Longitudinal alignment of the undulator pair with
respect to beam line elements was done by an aluminium profile.
Angular misalignments can cause contributions to the oscillation
length, so stiffness against torsion or bending of the setup while
moving the undulator was crucial. A welded steel construction
supported the setup.

%--------------------------------------------------------------------- 
\subsection{Undulator fields}
%--------------------------------------------------------------------- 

\begin{figure}[tbp]
  \centering
  \includegraphics[width=\textwidth]{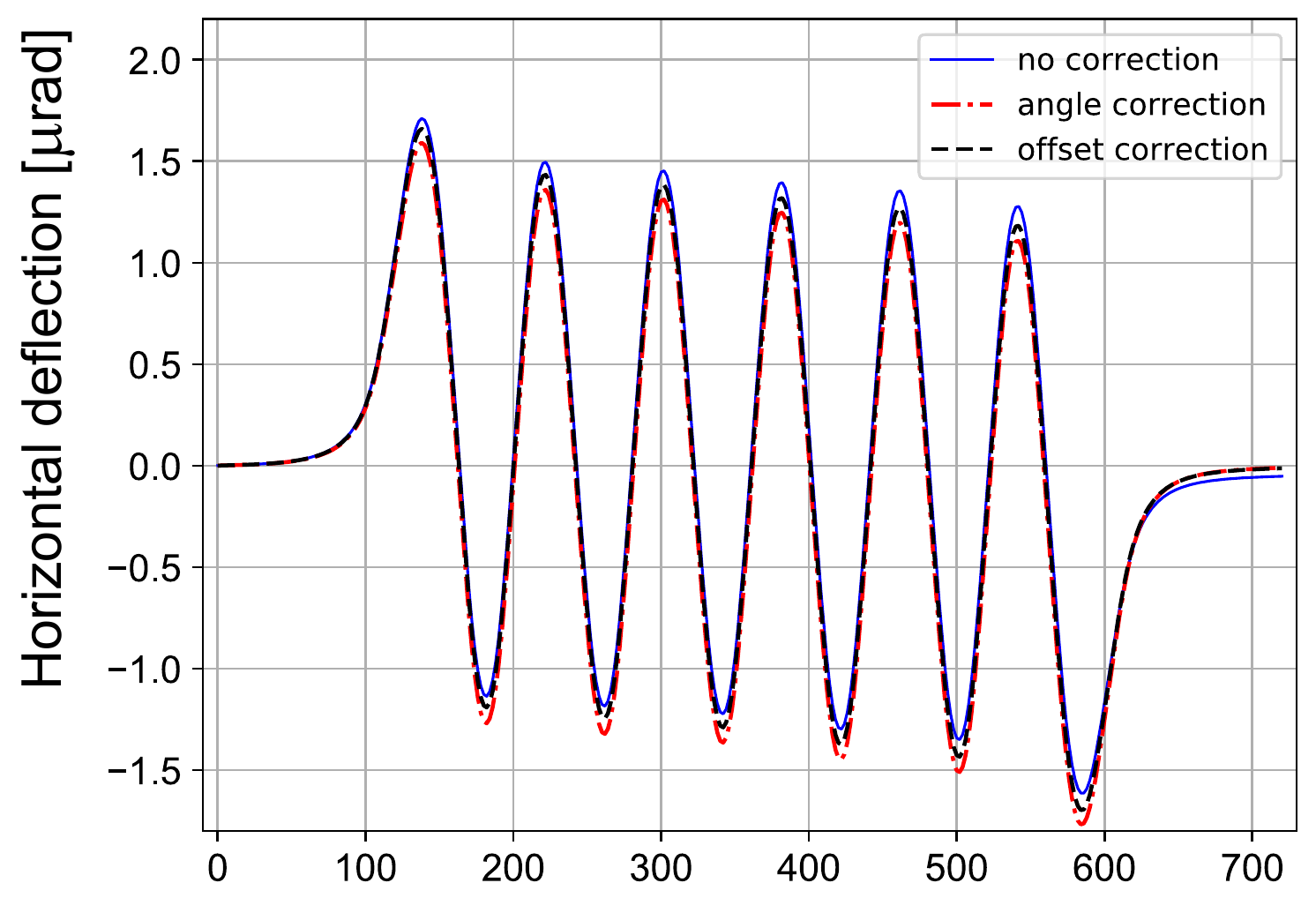}
  \includegraphics[width=\textwidth]{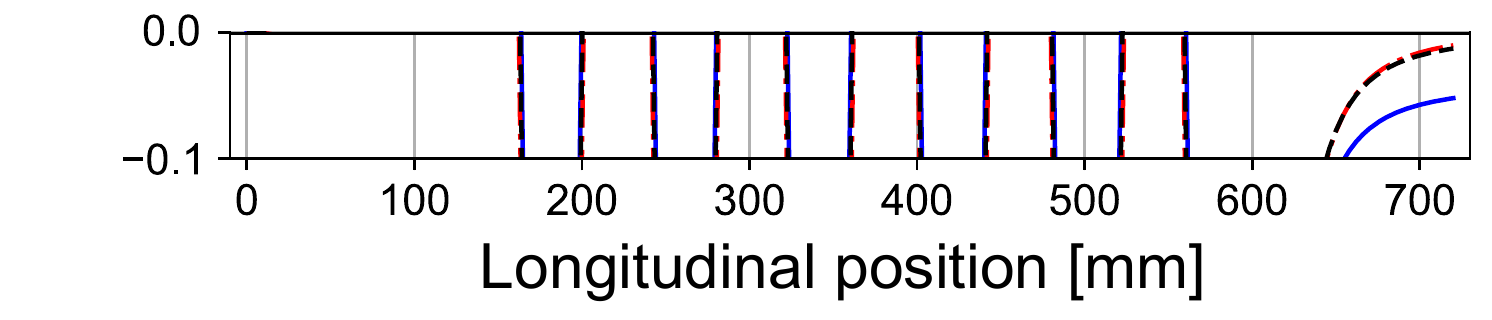}
  \caption{First integrals along the $z$-direction of the measured
    magnetic field component in the $y$-direction inside one of the
    undulators for $B_0 = 0.1$\,T. They represent the deflection angle
    $\zeta(z)$ of the electron beam in the $x$-direction Top:
    Integrals before (solid blue line), after compensation of the
    deflection angle (dot-dashed red line), and after compensation of
    the offset (dashed black line) are discussed in the text. Bottom:
    The close-up visualizes the residual angle at $z = 700$\,mm with
    and without adjustments of $|\zeta_{\text{res}}| < 10$\,$\mu$rad
    and $\zeta_\text{{res}} \approx -50$\,$\mu$rad, respectively.}
  \label{fig:FirstIntegrals}
\end{figure}

\begin{figure}[tbp]
  \centering
  \includegraphics[width=\textwidth]{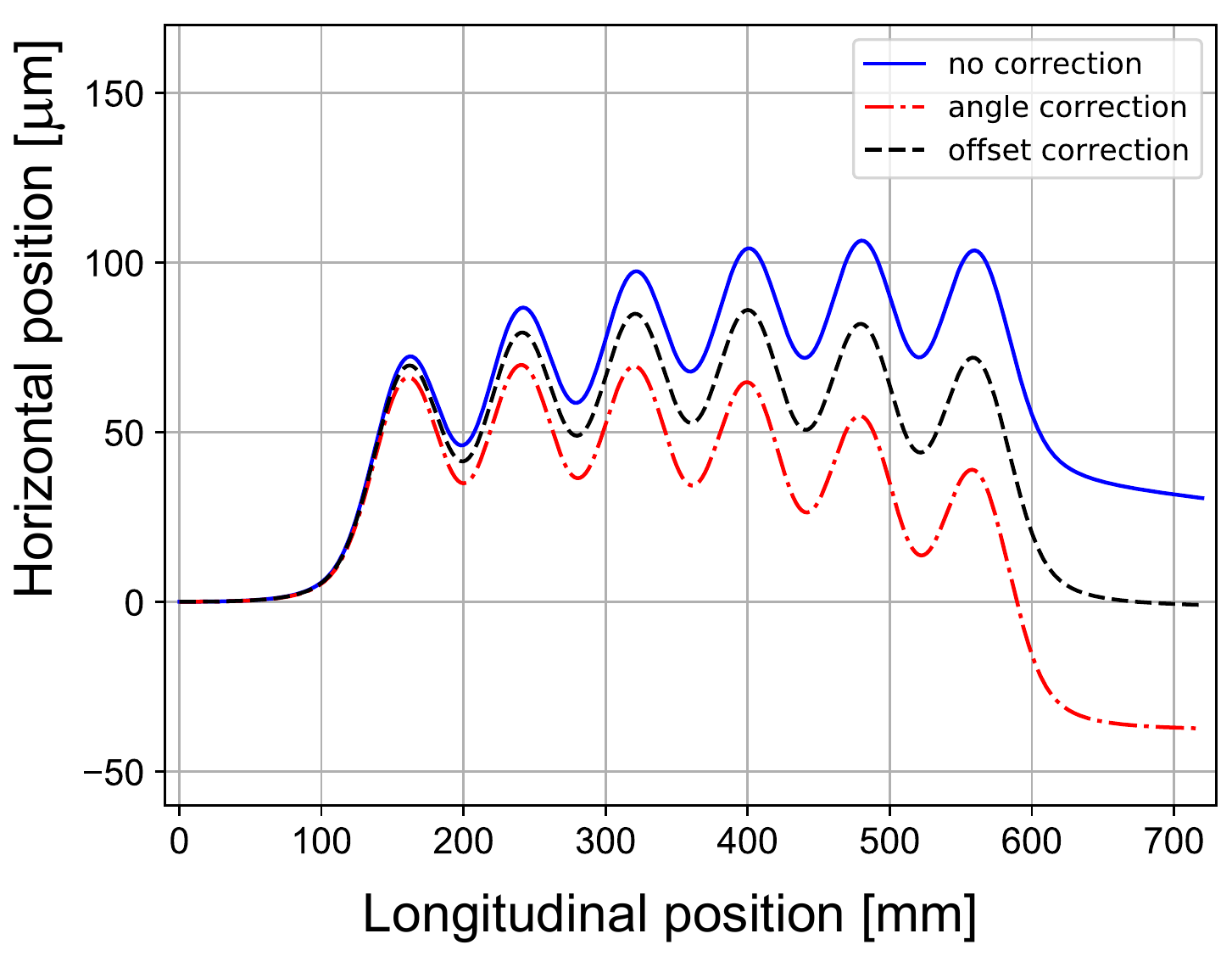}
  \caption{Second integrals along the $z$-direction of the measured
    magnetic field component in the $y$-direction inside one of the
    undulators for $B_0 = 0.1$\,T. They represent the offset $x(z)$ of
    the electron beam in $x$-direction. The integrals before (solid
    blue line), after the compensation of the deflection angle
    (dot-dashed red line), and after the compensation of the offset
    (dashed black line) are discussed in the text. The residual offset
    at $z = 700$\,mm with and without coil current adjustments was
    $|x_{\text{res}}| < 2$\,$\mu$m and $x_{\text{res}} \approx
    30$\,$\mu$m, respectively.}
  \label{fig:SecondIntegrals}
\end{figure}

To minimize systematic uncertainties in the beam energy measurements,
the beam inside of the two undulators should follow identical
trajectories and the light emission cones from the undulators should
be on-axis. The beam deflection angle is proportional to the integral
along the $z$-direction over the magnetic field projected on the beam
momentum:
\begin{align}
  \zeta(z) = \frac{e}{m_e \gamma c} \int_0^z\! B_y(z')\, dz'\,.
\end{align}
The second integral over the vertical magnetic field component
determines the spatial offset of the beam in horizontal direction:
\begin{align}
  x(z) = \frac{e}{m_e \gamma c} \int_0^z\! \int_0^{z'}\! B_y(z'')\, dz''
  dz'\,.
\end{align}

A Hall probe~\cite{Group3:2006} was moved along the beam axis to
generate a linear map of the magnetic
fields. Fig.~\ref{fig:FirstIntegrals} shows the stepwise computed
deflection angle for three configurations of the coil
currents. Without any adjustments to the current distributions in the
undulators, a residual angle $\zeta_{\text{res}} \approx
-50$\,$\mu$rad remained at $z = 700$\,mm, where the beam leaves the
field region. It was minimized by increasing the current $I_1$ in the
first pair of coils and decreasing $I_{13}$ in the last pair by a
smaller amount. The residual deflection angle after this adjustment
was $|\zeta| < 10$\,$\mu$rad.  The compensation of the deflection
angle led to a residual beam offset of $x_{\text{res}} \approx
-40\,\mu$m, being of the same size as for the initial current
distribution but of opposite sign.  This offset was subsequently
corrected by changing the coil currents symmetrically according to the
condition $\Delta I_1 = - \Delta I_{13}$, where $\Delta I_1$ denotes
the change of current in the first coil pair and $\Delta I_{13}$ in
the last coil pair. This condition ensured that the residual
deflection angle remained constant and close to
zero. Fig.~\ref{fig:SecondIntegrals} shows the second integrals along
the $z$-direction representing the beam trajectories in the horizontal
direction for the three coil current configurations.  Without coil
current adjustment, the residual offset was $x_{\text{res}} \approx
30$\,$\mu$m and after both adjustments, the residual offset was
$|x_{\text{res}}| < 2$\,$\mu$m. This two-step correction procedure was
successfully performed for both undulators. The final coil current
configuration ensured the emission of synchrotron light from identical
electron beam trajectories up to the level of the accuracy of the hall
probe.

%--------------------------------------------------------------------- 
\subsection{Optical interferometer system}
%--------------------------------------------------------------------- 

\begin{figure}[tbp]
  \centering
  \includegraphics[width=\textwidth]{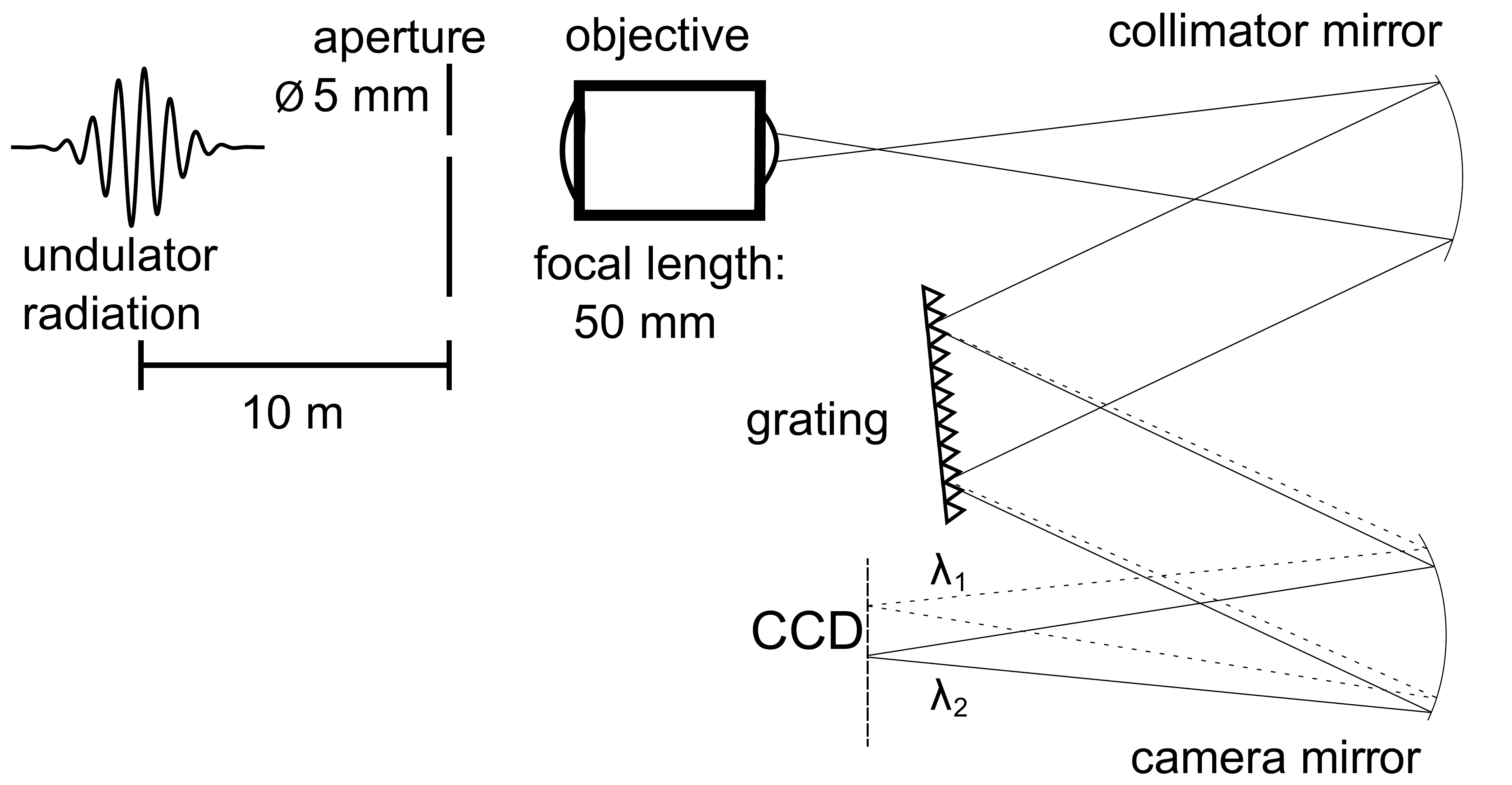}
  \caption{Schematic drawing of the optical spectrometer (not to
    scale). The setup consisted of a high-resolving Czerny Turner
    monochromator comprising two spherical mirrors and one diffraction
    grating in between. The light beam entered from the top left and
    was collimated by an aperture positioned in front of the imaging
    objective. The objective projected the beam to the focal point of
    the collimator mirror, so that a parallel beam was reflected onto
    the planar grating. The reflected and diffracted light was focused
    by the camera mirror into a focal plane where a CCD camera was
    placed.}
  \label{fig:SchematicMonochromator}
\end{figure}

The synchrotron light was observed with an optical interferometer
system based on a Czerny Turner monochromator of type {\tt HR 640}
from Jobin Yvon~\cite{Horiba:1993} at a distance $L \approx 10$\,m
from the undulators.  A schematic drawing of the spectrometer is shown
in Fig.~\ref{fig:SchematicMonochromator}.  A 5\,mm diameter aperture
in the light path collimated the synchrotron light to an angular
acceptance of $\Delta \theta = \pm 0.25$\,mrad. This beam was
incident on an imaging objective with 50\,mm focal length.  The
objective projected the beam to the focal point of the collimator
mirror.  The beam was then reflected by the mirror, which rendered it
parallel and directed it to the planar grating. The monochromator
could be equipped with a prism of low dispersion and large wavelength
acceptance or, alternatively, with a grating of 1200\,lines per
mm. For divergent light, the grating provided a resolution of $\delta
\lambda/\lambda = 3 \times 10^{-5}$. The monochromator directed the
dispersed light spectrum to the camera mirror, which focused the image
of the vertical entrance aperture on the focal plane. The components
of the optical system were aligned with a collimated and
frequency-doubled Nd:Yag laser at 532\,nm.

For the calibration of absolute wavelengths and of the dispersion of
the monochromator, optical light sources with well known spectral
lines were used. For the spectral range from 396 to 410\,nm, the two
Hg lines at 404.6565\,nm and 407.7837\,nm ($\delta \lambda \approx
10^{-4}$\,nm)~\cite{NIST:ASD:2016} of a mercury vapor lamp were best
suited and the systematic errors from the absolute spectral
calibration were negligible.

%--------------------------------------------------------------------- 
\subsection{CCD camera}
%--------------------------------------------------------------------- 

\begin{figure}[tbp]
  \centering
  \includegraphics[width=0.8\textwidth]{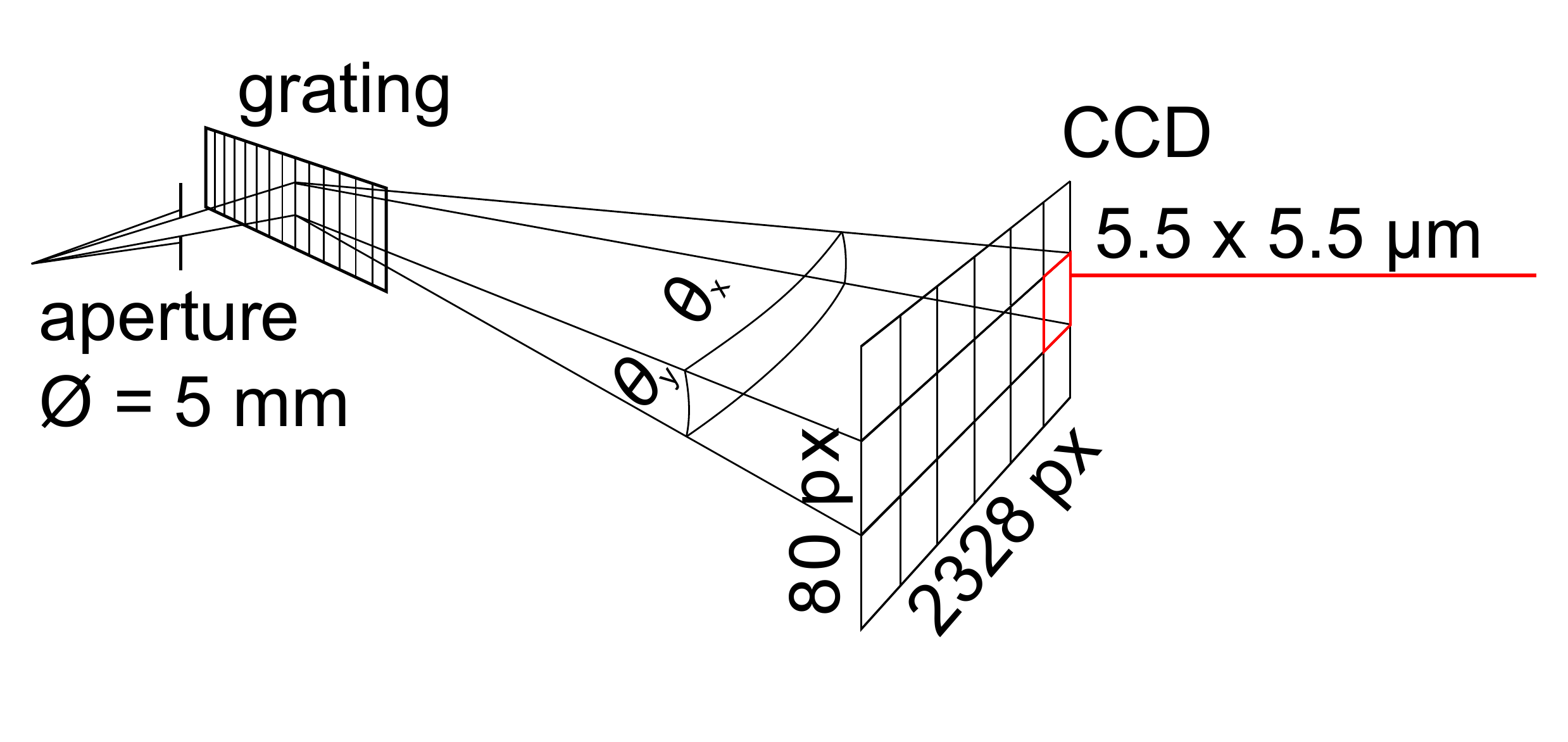}
  \caption{Schematic drawing of the CCD setup (not to scale). The
    2\,328 rows of the CCD resolved the light spectrum in the
    dispersive direction. An image of the entrance aperture could be
    observed in the non-dispersive direction with 80 of the 1\,750
    columns of the CCD.}
  \label{fig:SchematicCCD}
\end{figure}

A camera of type {\tt avA2300-25gm} from Basler~\cite{Basler:2011}
with a charged-coupled device (CCD) image sensor of $2\,328 \times
1\,750$ pixels (px) from Kodak~\cite{KAI:2012} was used as optical
detector in the interferometer system.  It can deliver 26 frames per
second with 4 megapixels per frame. This camera was chosen for its
high quantum efficiency $\eta >$ 30\,\% at 400\,nm in the short
wavelength band. Its pixel size of $5.5 \times 5.5\,\mu$m$^2$ matched
the spectrometer resolution. The movement of the undulator was
synchronized with the image taking of the camera so that each image
corresponded to a fixed distance between the undulators.

Fig.~\ref{fig:SchematicCCD} shows a schematic drawing of the CCD
setup.  The chip size in the horizontal direction could fully be used
in the focal plane of the spectrometer. In the vertical direction, the
light cone leaving the spectrometer was projected onto 80 pixels.  In
the measurements that used a grating, the high dispersion and the
limited acceptance led to a light intensity not exceeding a few
thousand photons per second and pixel for electron beam currents of
$I_{\text{beam}} \approx 1\,\mu$A.  Typical exposure times were 2.5\,s
long, which is the maximum exposure time for this camera. Four pixels
in the vertical direction were binned together to compensate for the
low light intensity.  Further minimization of the relative counting
error was achieved by averaging 5 consecutive images.  Each image has
been taken with a gain setting of 600 using the full ADC depth of
12\,bit and has been stored in a 16\,bit image file.

%--------------------------------------------------------------------- 
\section{Synchrotron radiation measurements at MAMI}
\label{sec:images}
%--------------------------------------------------------------------- 

\begin{figure}[tbp]
  \centering
  \includegraphics[height=.8\textheight]{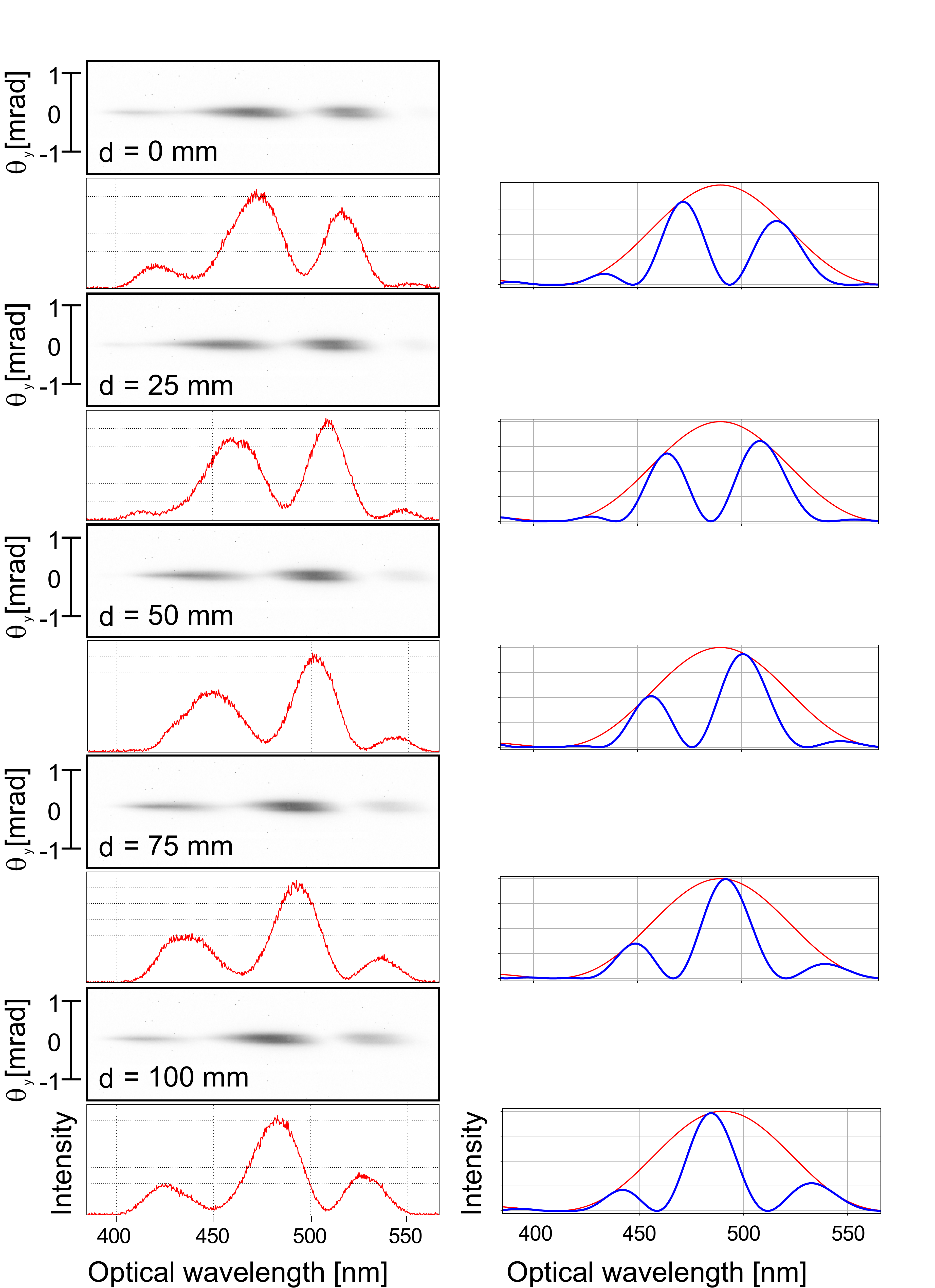}
  \caption{Evolution of the synchrotron spectrum taken with a prism as
    a monochromator for a variation of the undulator distance $d$ from
    0 to 100\,mm. Left: The 16-bit grayscale inverted images show
    areas of 600\,px $\times$ 160\,px.  The rows (horizontal
    direction) resolved the wavelength and the columns (vertical
    direction) provided an image of the entrance aperture in the
    non-dispersive direction. The spectra below the images show the
    intensity distributions for the single row at minimum
    $\theta_y$. Right: Analytical prediction of the corresponding
    spectra assuming perfect optical imaging and fully coherent
    synchrotron radiation. The red lines show the spectral envelopes.}
  \label{fig:CCDSpectraPrism}
\end{figure}

\begin{figure}[tbp]
  \centering
  \includegraphics[width=\textwidth]{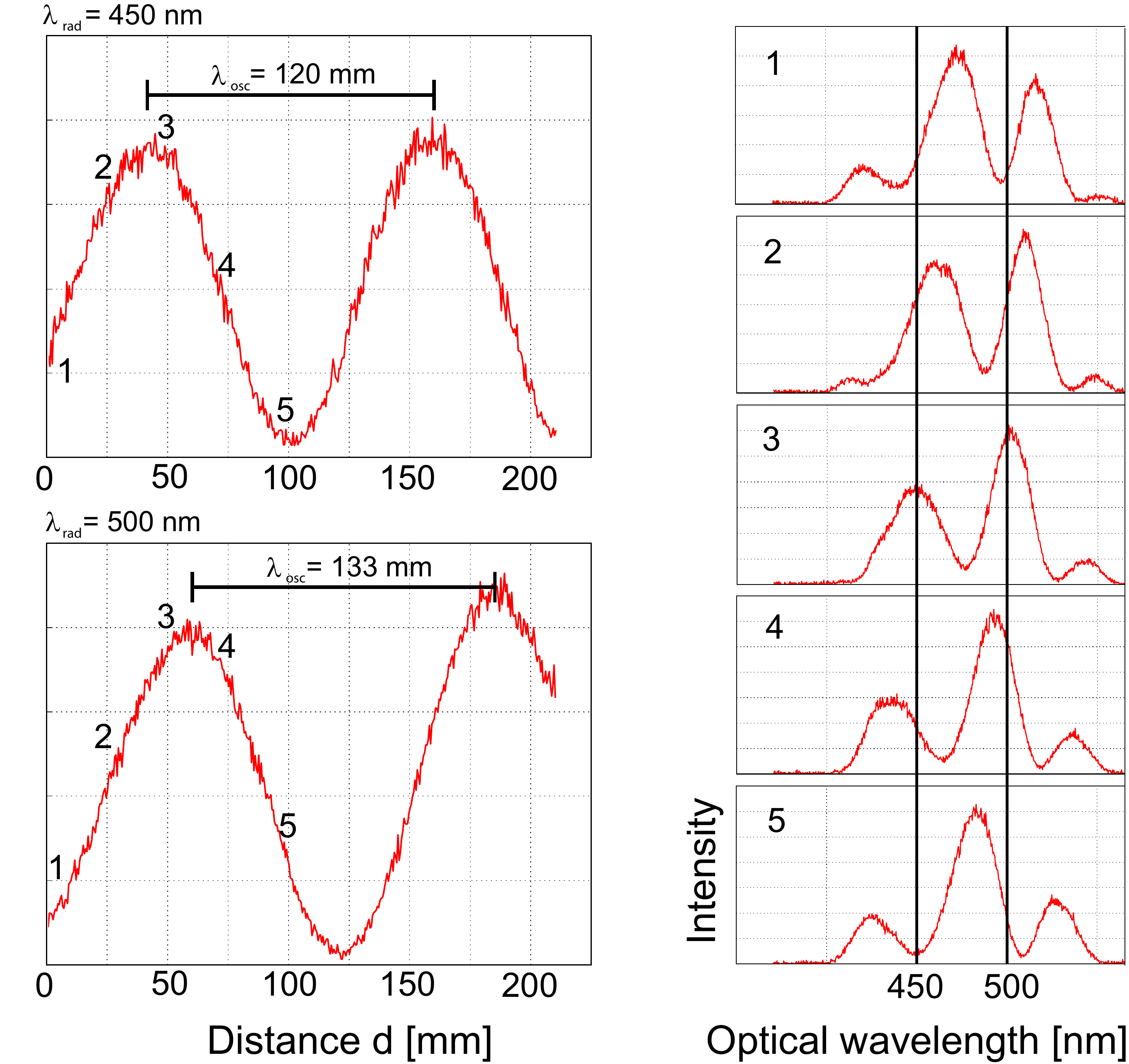}
  \caption{Right: Evolution of the intensity distributions for the
    single row at minimum $\theta_y$ with the distance $d$. Left:
    Intensities in two pixels of one CCD row for two selected
    wavelengths at 450\,nm and 500\,nm. The numbers describe the
    correspondence between the spectra on the right and the
    interference oscillations on the left.  The oscillation length
    increased by approximately 11\,\% when selecting a $500/450
    \approx 1.11$ larger wavelength. The beam energy was approximately
    195\,MeV.}
  \label{fig:IntensityEvolutionPrism}
\end{figure}

\begin{figure}[tbp]
  \centering
  \includegraphics[height=.8\textheight]{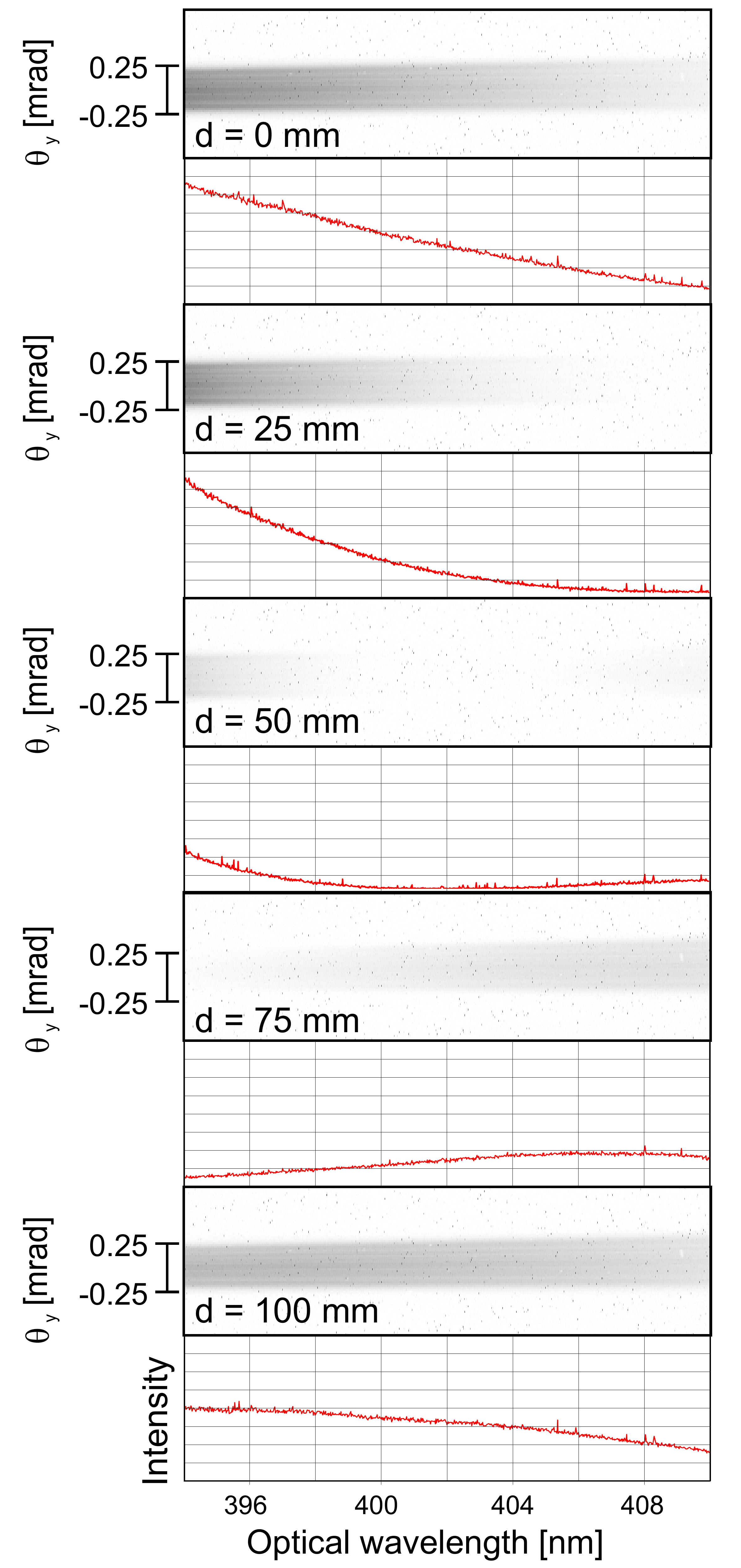}
  \caption{Evolution of the synchrotron spectrum taken with a grating
    as a monochromator for a variation of the undulator distance $d$
    from 0 to 100\,mm. Compared to Fig.~\ref{fig:CCDSpectraPrism}, the
    grating provided an approximately 60 times larger dispersion.  The
    16-bit grayscale inverted images show areas of 2\,328\,px $\times$
    320\,px with four CCD pixels combined to one in the vertical
    direction.}
  \label{fig:CCDSpectraMonochromator}
\end{figure}

The two undulators were used as synchrotron radiation sources at MAMI
in two consecutive beam-times to verify their proposed application for
absolute beam energy measurements. The nominal beam energy was
195.2\,MeV as determined by the standard method introduced in
Section~\ref{sec:MAMI}.  The undulator pair acted as a source of
coherent synchrotron light emitted under very small angles with
respect to the electron beam and leaving the beam pipe through a
window. The first beam-time used a prism as a monochromator. It served
as a demonstration of the functionality of the experimental setup, of
the coherence of the synchrotron radiation, and of the precision of
the optical system.  Fig.~\ref{fig:CCDSpectraPrism} shows a series of
images and corresponding synchrotron spectra taken for five different
undulator distances in steps of $25\,$mm. Non-linearities of the
prism, aberrations of the optics in front of the CCD, and incomplete
coherence of the undulator radiation led to small differences between
the predicted and measured spectra.

For the determination of the oscillation length the distance was
varied in steps of 0.5\,mm so that approximately 200 separate spectra
could be analyzed.  The evolution of the intensity distribution is
shown in Fig.~\ref{fig:IntensityEvolutionPrism} for the five selected
positions together with the full intensity variation for two selected
wavelengths near 450\,nm and 500\,nm.  As expected, the intensity for
each wavelength in the spectrum varied periodically as the undulator
moved.  It was verified that the oscillation length increased
proportional to the wavelength. The right column of
Fig.~\ref{fig:CCDSpectraPrism} shows the analytical predictions
assuming perfect optical imaging and no coherence loss.

The on-axis coherence $C$ for perfectly aligned undulators is limited
only by the beam emittance. For a beam energy of 195\,MeV from the
RTM3, the normalized beam emittance in the horizontal plane and
vertical plane is $\epsilon_x = 4\pi$\,$\mu$m\,mrad and $\epsilon_y =
1\pi$\,$\mu$m\,mrad, respectively~\cite{Euteneuer:1994}. In addition,
the coherence decreases with non-vanishing observation angles
$\theta_{x,y}$.  The conditions to be satisfied in order to avoid a
decrease of on-axis coherence $C$ larger than $1/\sqrt{2}$ for a
selected wavelength $\lambda$ and distance $L$ between detector and
undulators are~\cite{Dambach:1998}:
\begin{align}
  \frac{d + L_U}{L} \frac{\epsilon_{x,y}}{\pi} \leqslant
  \frac{\lambda}{4\pi} \qquad \text{and}\qquad \theta_{x,y}\ (d+L_U)
  \sqrt{\frac{\epsilon_{x,y}}{\pi L}} \leqslant
  \frac{\lambda}{2\pi}\,.
  \label{eq:Coherence}
\end{align}
For a wavelength $\lambda = 400$\,nm, a distance $L = 10$\,m and the
given emittances, the first coherence condition requires $d \leqslant
79$\,m, which is fulfilled by the extends of the experimental
setup. The second condition sets a limit to the observation angles of
$\theta_{x} \leqslant 3.2$\,mrad and $\theta_{y} \leqslant 6.4$\,mrad,
which is fulfilled by an alignment of the setup with sub-mrad angular
precision.

In the second beam-time, the measurements have been continued with a
grating monochromator that provided an approximately 60 times larger
dispersion.  The horizontal axis of the camera, aligned in the
dispersive direction of the spectrometer, resolved the wavelengths
with 162\,px$/$nm. The chip size with all 2\,328 pixels in a row
covered a range of approximately 15\,nm.  The spectrum was spread
vertically over 80 pixels, with each point determined by the sum of
four pixels. A single row had an angular acceptance of $\Delta
\theta_y = \pm 0.025$\,mrad. Fig.~\ref{fig:CCDSpectraMonochromator}
shows a series of camera images for different distances $d$.  The
spectra below the images show the intensity distributions for the
single row at minimum $\theta_y$.  For a step width of 1\,mm between
undulator positions, the duration for one measurement run was 1 hour.

%--------------------------------------------------------------------- 
\section{Determination of MAMI beam energies}
\label{sec:beamenergy}
%--------------------------------------------------------------------- 

\begin{figure}[tbp]
  \centering
  \includegraphics[width=0.7\textwidth]{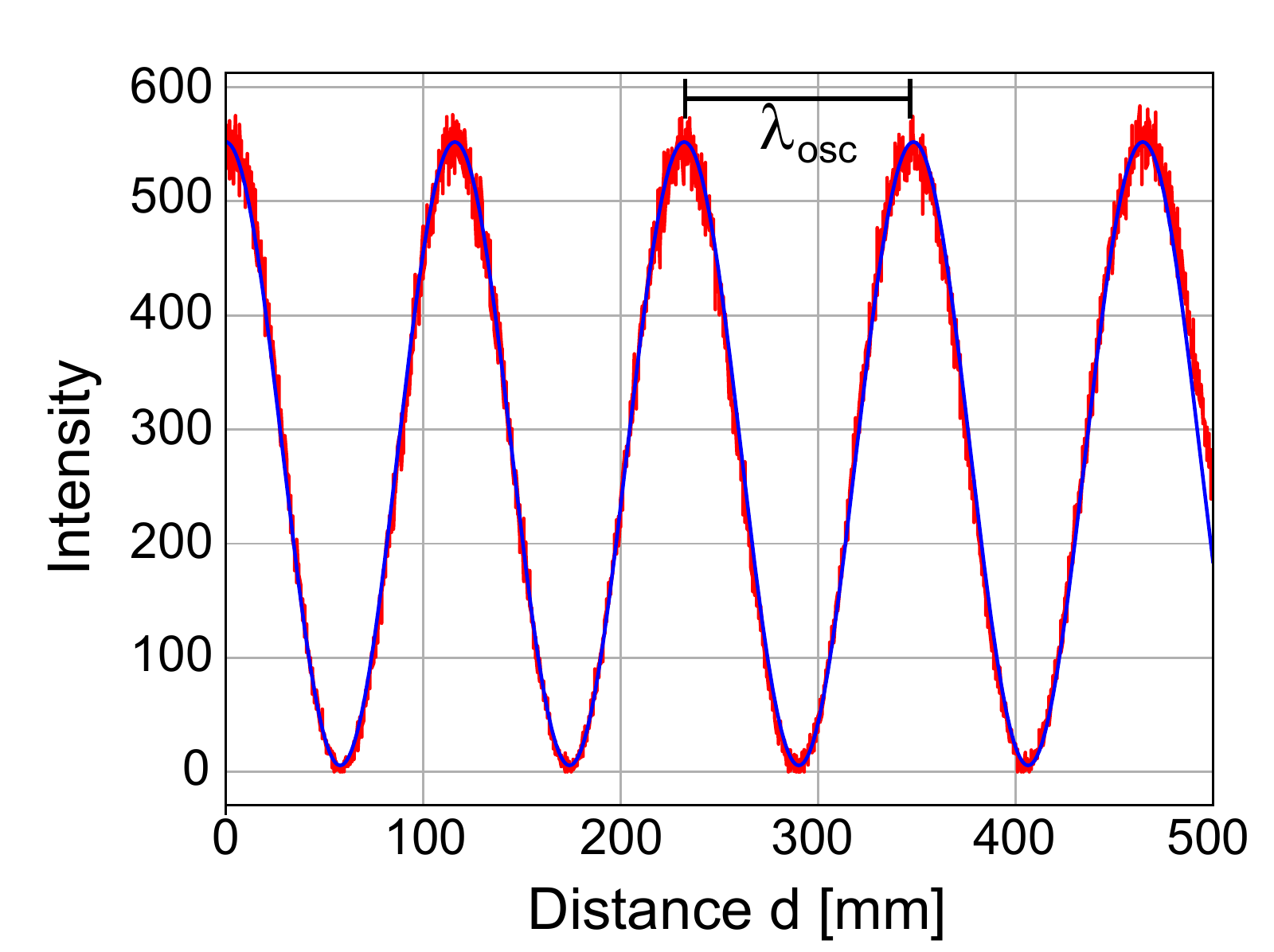}
  \includegraphics[width=0.7\textwidth]{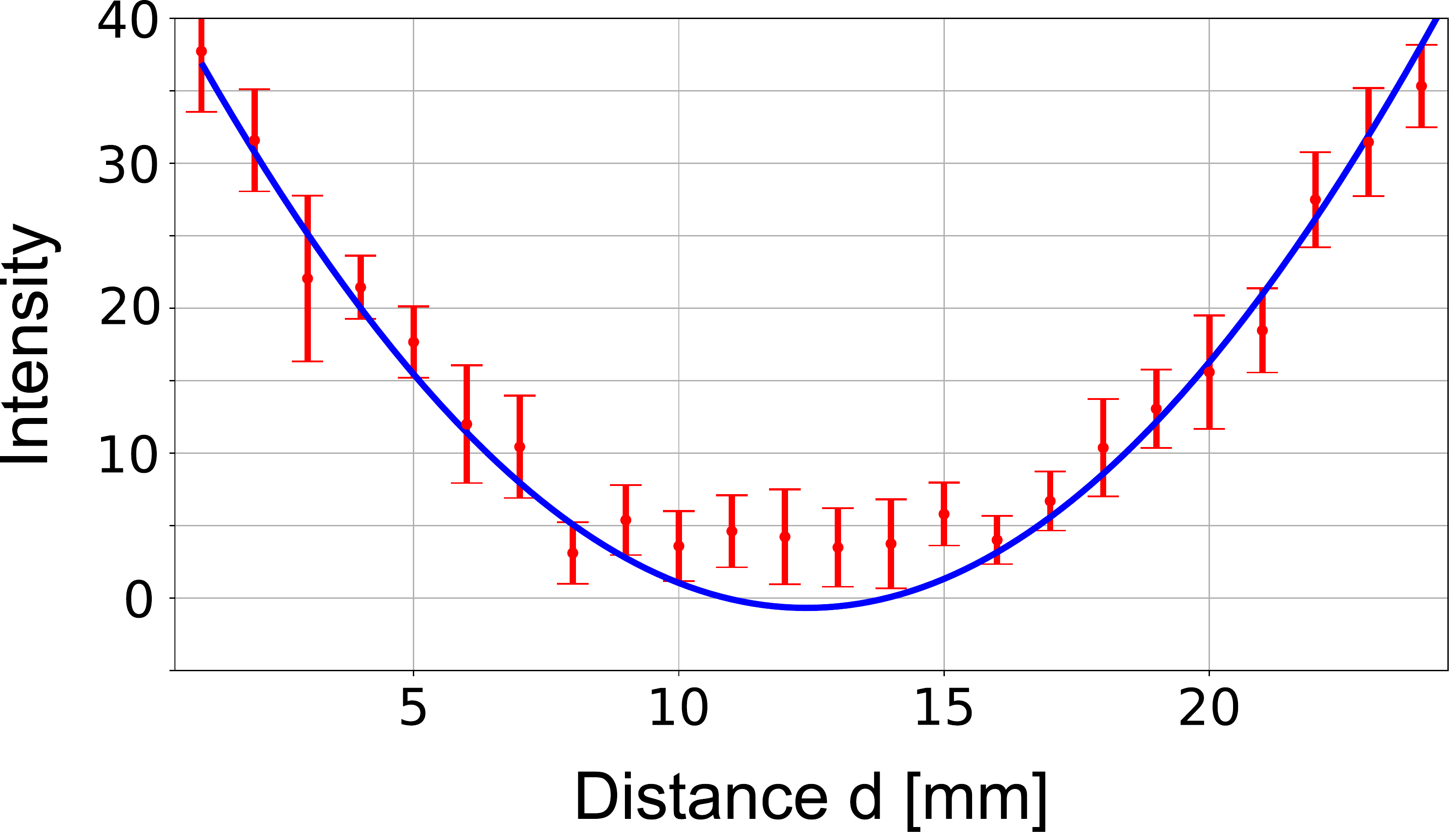}
  \caption{Top: Intensity as a function of distance $d$ taken from a
    single CCD pixel corresponding to one exemplary wavelength band at
    $\lambda_{\text{rad}} \approx 396$\,nm and minimum $\theta_y$ (red
    data points). The data was fit with a sine function (blue line)
    from which an oscillation length $\lambda_{\text{osc}}=
    116.12$\,mm was determined.  The nominal beam energy was
    195.2\,MeV. The fit function starts to deviate from the data at
    distances $d \gtrsim 450$\,mm. Bottom: Data (red points) and fit
    function (blue line) near the first minimum, where systematic
    deviations were observed.}
  \label{fig:IntensityEvolution396nm}
\end{figure}

\begin{figure}[tbp]
  \centering
  \includegraphics[width=\textwidth]{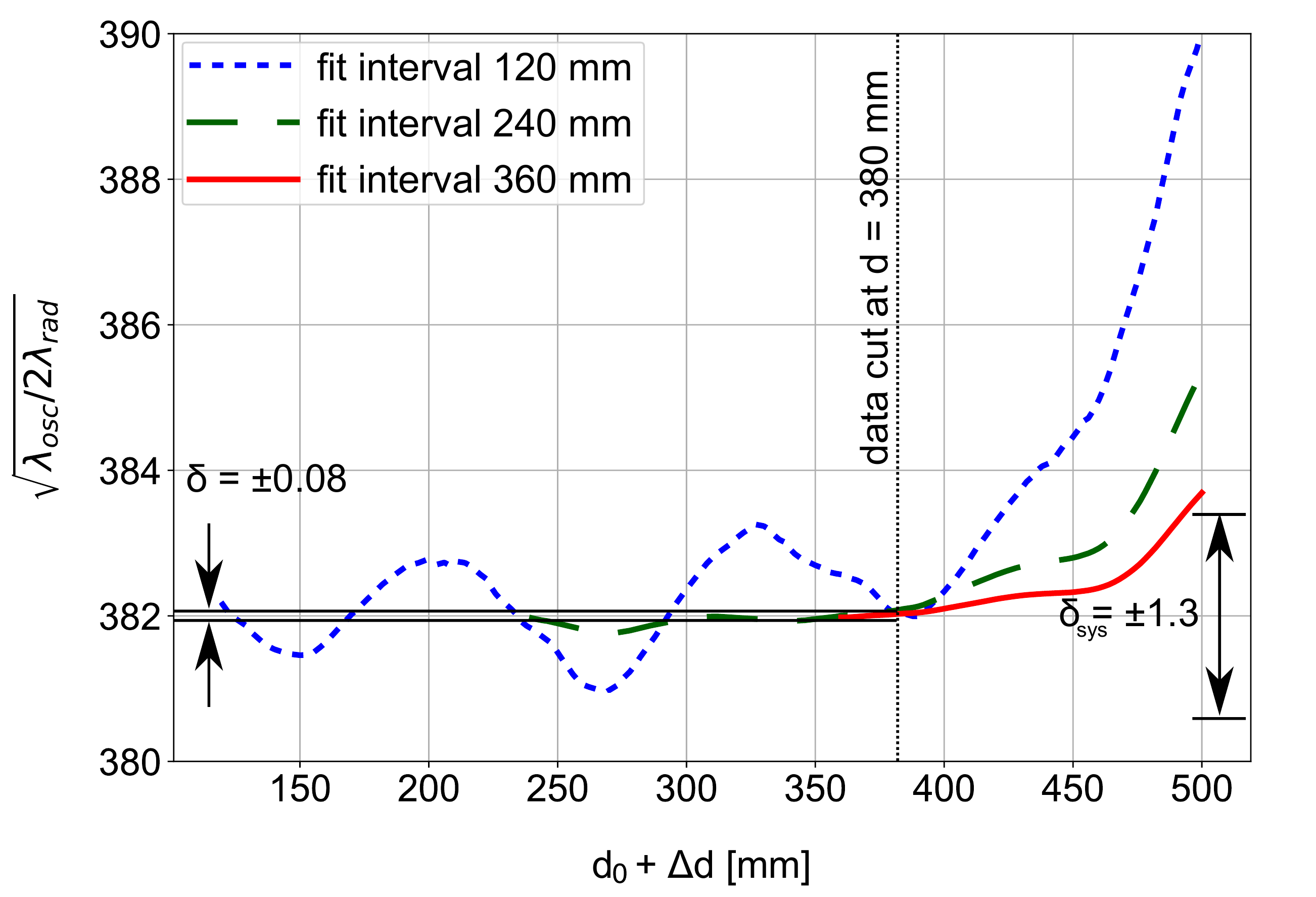}
  \caption{Interference observable $\tilde\gamma =
    \sqrt{\lambda_{\text{osc}}/2\lambda_{\text{rad}}}$ averaged over
    2\,328 wavelength bands. The values were deduced from the
    interference oscillations during one measurement run for different
    fit intervals $\Delta d$ between 120 (blue dots) and 360\,mm (red
    line), corresponding to 1 to 3 oscillation lengths, by varying the
    start point $d_0$ of the fit interval. The total systematic
    uncertainty in the interference observable of $\delta \tilde
    \gamma = \pm 1.3\ {\text{(sys.)}}$ as well as a data cut at $d =
    380$\,mm are indicated. The nominal beam energy was $195.2 \pm
    0.3$\,MeV, corresponding to a Lorentz factor $\gamma_{\text{nom}}
    = 382.0 \pm 0.6$.}
  \label{fig:gammaSystematicErrors}
\end{figure}

\begin{figure}[tbp]
  \centering
  \includegraphics[width=\textwidth]{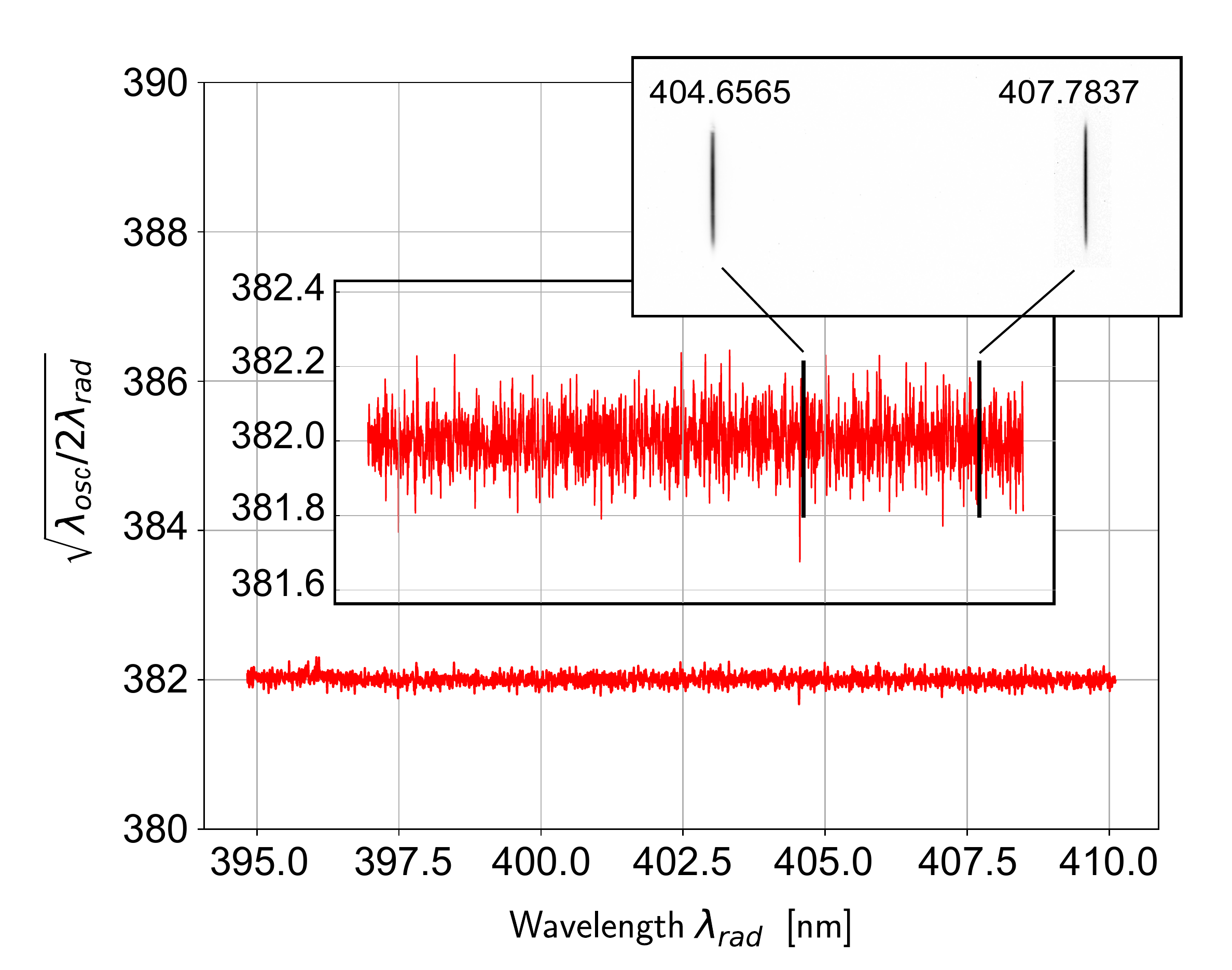}
  \caption{Interference observable $\tilde\gamma =
    \sqrt{\lambda_{\text{osc}}/2\lambda_{\text{rad}}}$ as
    simultaneously observed for 2\,328 different wavelength bands at
    $|\theta_y| < 0.025$\,mrad. The values were deduced from the
    interference oscillations during one measurement run for distances
    $d$ from 0 to 380\,mm. The inset is a close-up view so that the
    point-to-point fluctuations in the data are visible. No wavelength
    dependence was observed. The interference observable
    $\tilde\gamma$ represents the Lorentz factor $\gamma$ of the
    electron beam plus off-axis contributions. The optical
    interferometer system has been calibrated with Hg emission lines
    at 404.6565 and 407.7837\,nm, whose positions are indicated by the
    vertical markers. The precision of the calibration is visualized
    by the CCD image. One CCD pixel provided a wavelength bandpass of
    $\Delta \lambda_{\text{rad}} = 6.6 \times 10^{-3}$\,nm. The
    nominal beam energy was $195.2 \pm 0.3$\,MeV, corresponding to a
    Lorentz factor $\gamma_{\text{nom}} = 382.0 \pm 0.6$.}
  \label{fig:gammaDetermination}
\end{figure}

Fig.~\ref{fig:IntensityEvolution396nm} shows the periodic variation of
the intensity at one exemplary wavelength band at
$\lambda_{\text{rad}}=$ 396\,nm and at minimum $\theta_y$ as a
function of the distance $d$. Based on the least squares method, the
free parameters intensity $I_0$, oscillation length
$\lambda_{\text{osc}}$, oscillation phase $\phi_{\text{osc}}$, and
intensity offset $I_{\text{offset}}$ were determined by a fit of the
data to the following function:
\begin{align}
  I(d) = I_0 \sin \left( 2\pi \frac{d}{\lambda_{\text{osc}}} +
    \phi_{\text{osc}} \right) + I_{\text{offset}}\,.
  \label{eq:fitfunc}
\end{align}

For the determination of the electron beam energy, the single row of
the CCD image at $|\theta_y| < 0.025$\,mrad was used, which defined the
vertical center of the synchrotron radiation cone.  From the intensity
oscillation, the interference observable $\tilde\gamma =
\sqrt{\lambda_{\text{osc}}/2\lambda_{\text{rad}}}$ could be extracted
for each wavelength band corresponding to a single CCD pixel. This
observable represents the Lorentz factor $\gamma$ of the electron beam
plus contributions depending on differences between light emission and
observation angles. With the accepted wavelength band of the
monochromator covering all 2\,328 CCD pixels in the horizontal
direction of the camera, the same number of simultaneous
determinations of interference oscillations could be performed in one
measurement run.

\begin{figure}[tbp]
  \centering
  \includegraphics[width=\textwidth]{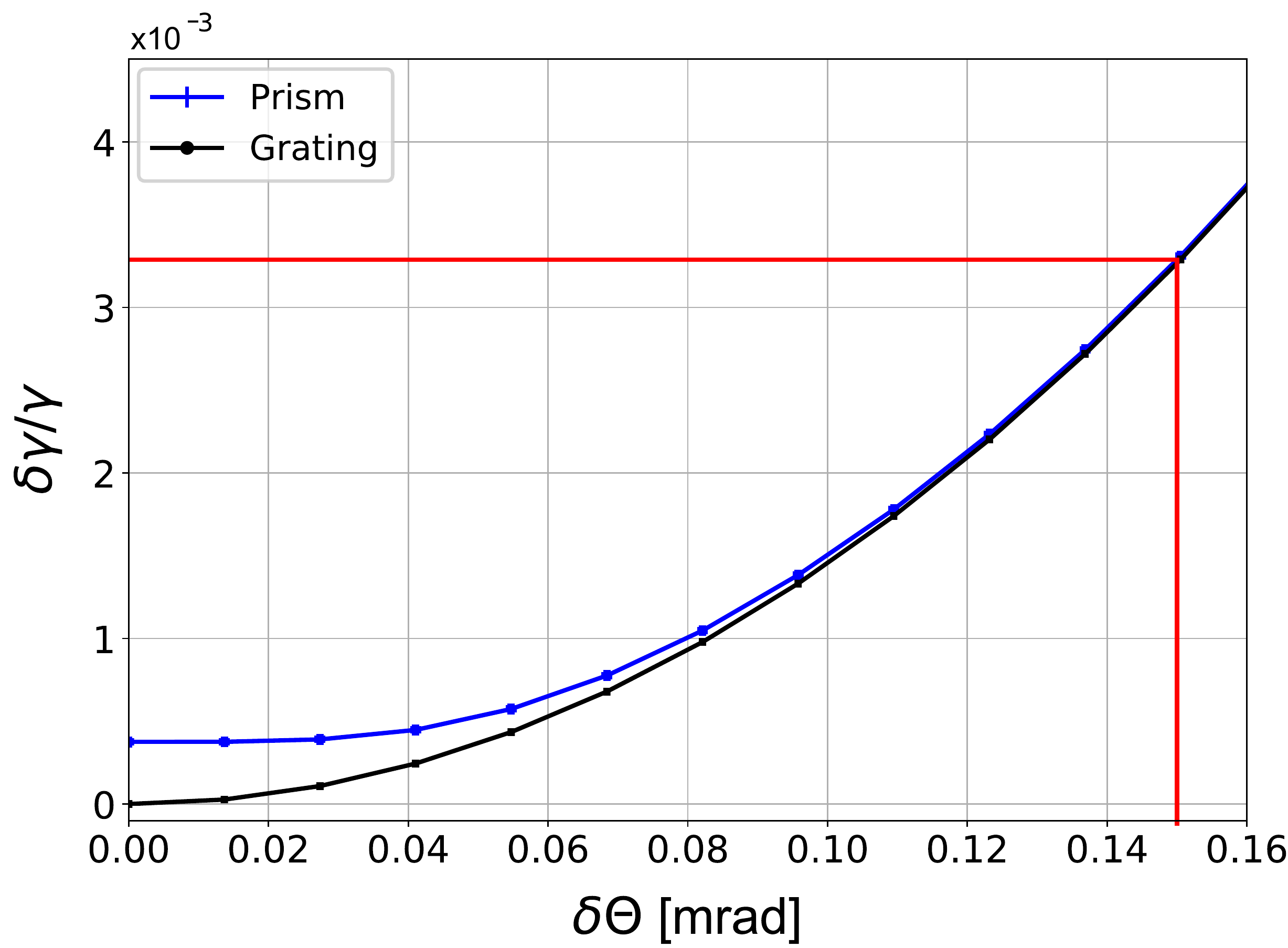}
  \caption{Relative uncertainty on the measured Lorentz factor
    $\gamma$ of the beam as a function of the uncertainty on the
    observation angle $\delta \theta$ for the two types of
    monochromators used in the beam-times, a low-dispersion prism and
    a high-dispersion grating. For the setup with prism, uncertainties
    in the calibration, non-linearities, and optical aberrations
    contributed to the uncertainty. A conservative estimate for the
    angular uncertainty of $\delta \theta_x \approx 0.15$\,mrad
    resulted in a systematic uncertainty of $\delta \gamma\
    (\text{sys.}) = \pm $ 1.3.}
  \label{fig:deltagamma}
\end{figure}

The analysis of the residua between the fit function and data revealed
some systematic deviations, especially for the largest scanned
distances of $d \gtrsim 450$\,mm and near the positions of minimum
intensity. Fig.~\ref{fig:IntensityEvolution396nm} illustrates such
deviations for one exemplary wavelength band. The robustness of the
fitting procedure was then studied by varying the length of the fit
interval $\Delta d$ and the start point $d_0$ of the fit
interval. Fig.~\ref{fig:gammaSystematicErrors} shows the interference
observable $\tilde\gamma =
\sqrt{\lambda_{\text{osc}}/2\lambda_{\text{rad}}} \approx 382$
averaged over all wavelength bands. For $\Delta d$ approximately equal
to one oscillation length, systematic variations on the order of
$10^{-3}$ with a periodicity of one oscillation length were
observed. These systematic variations decreased rapidly when the fit
interval was increased. For the final data set, the residua between
the fit function and data points followed a normal distribution with a
width of $\sigma = 0.08$.

A Monte Carlo simulation for the light intensity as a function of $d$
was performed using a sine function with the nominal value for the
oscillation length. The simulated data followed a normal distribution
with respect to the sine function for intensities above 10\,\% of the
maximum intensity and a Poissonian distribution for intensities
below. This approach ensured the positiveness of intensity values as
it is the case in real cameras. Fits to the simulated data with an
sine function produced alternating deviations on the order of
$10^{-3}$ with a periodicity similar to the oscillation length,
revealing that the observed systematic deviations were caused by the
camera. The simulation confirmed that the effect on the extracted fit
parameters becomes negligible, when longer fit intervals are used.

At larger distances $d \gtrsim 450$\,mm, the permanent excursion of
the interference observable above the systematic error band indicated
the onset of additional contributions. The $\chi^2$ value, which was
evaluated with the standard procedure to provide a measure for the
goodness of the fit, also increased for larger start points $d_0$.
This behavior can be expected when additional contributions to the
sine function are present in the data. To minimize such contributions,
the data was cut at $d = 380$\,mm in all further analyzes.

The amplitudes of the intensity oscillations showed no significant
variation with increasing undulator distances.  The coherence $C$ of
the synchrotron light observed with the experimental setup was studied
by fixing the fit interval $\Delta d$ to one oscillation length. The
coherence as determined by $C = 2 I_0/(I_0 + I_{\text{offset}})$ was
$C > 0.99$ for $d_0$ from 0 to 300\,mm.

The relative uncertainties in the Lorentz factor $\gamma$ of the beam
were given by
\begin{align}
  \frac{\delta \gamma}{\gamma} = \frac{1}{2} \sqrt{\left( \frac{\delta
        \lambda}{\lambda} \right)^{\!2} + \left( \frac{\delta
        \lambda_{\text{osc}}}{\lambda_{\text{osc}}} \right)^{\!2}}
\end{align}
and were dominated by the uncertainty of the horizontal angle between
the electron beam and the observation axis of the synchrotron
radiation, that enters into the uncertainty of the oscillation length:
\begin{align}
  \frac{\delta \gamma}{\gamma} = \frac{1}{2} \sqrt{\left( \frac{\delta
        \lambda}{\lambda} \right)^{\!2} + \left( \frac{2 \gamma^2 \delta
        \theta^2}{1 + \gamma^2 \delta \theta^2} \right)^{\!2}}
\end{align}
Figure~\ref{fig:deltagamma} shows the relative uncertainty as a
function of the uncertainty on the observation angle $\delta \theta$
for the two types of monochromators used in the beam-times. For an
angular uncertainty of $\delta \theta_x \approx 0.15$\,mrad, the
resulting systematic uncertainty is $\delta \gamma\ (\text{sys.}) =
\pm $ 1.3. The uncertainty from the spectral calibration over the full
width of the focal plane led to a systematic uncertainty of at least
one order of magnitude smaller.  Since the statistical noise of the
CCD increased with intensity, the oscillation curves showed larger
fluctuations near the maximum positions as compared to the minimum
positions. The interference observable was calculated as the
arithmetic mean over all wavelength bands in the spectrum so that the
statistical error of a single measurement was reduced by a factor
$\sqrt{2\,328}$ and became negligible compared to systematic effects.

Fig.~\ref{fig:gammaDetermination} shows the measured interference
observable $\tilde \gamma =
\sqrt{\lambda_{\text{osc}}/2\lambda_{\text{rad}}}$ for the 2\,328
different wavelength bands taken during one measurement run. The
values were deduced from the observed interference oscillations for
$|\theta_y| < 0.025$\,mrad and the cut at $d = 380$\,mm was
applied. Deviations from a straight line would indicate systematic
effects possibly caused by alignment errors or imperfect tuning of the
undulator fields. No wavelength dependence was observed.  The
arithmetic mean value was corrected by a factor of $1.00152 \pm
0.00006$, which was determined from the integration over the finite
angular acceptance.

Most stable operation conditions for the accelerator can be achieved
after reaching a thermal equilibrium in all components, typically many
hours after the start-up. Because of the short beam-time, these
conditions may not have been reached during the measurements and the
nominal beam energy of 195.2\,MeV might have had a substantial
uncertainty of up to 0.3\,MeV corresponding to $\gamma_{\text{nom}} =
382.0 \pm 0.6$. The determined Lorentz factor $\gamma$ of the beam,
measured in 1 hour of data taking with the undulator setup, was:
\begin{displaymath}
  \gamma = 382.5876 \pm 0.0015\ (\text{stat.}) \pm 1.3\
  (\text{sys.})\,,
\end{displaymath}
where the systematic uncertainties were dominated by possible angular
misalignments. The corresponding beam energy
\begin{displaymath}
  E_\text{beam} = 195.5019 \pm 0.0008\ (\text{stat.}) \pm 0.7\
  (\text{sys.})\, \text{MeV}
\end{displaymath}
is consistent with the nominal beam energy.

%--------------------------------------------------------------------- 
\section{Summary and conclusions}
\label{sec:conclusion}
%--------------------------------------------------------------------- 

Relativistic electrons deflected by two identical magnetic sections
generate interfering synchrotron radiation that can be used for the
precise diagnostics of the electron beam energy when the distance
between the sources is varied. A pioneering experiment has been
carried out at MAMI to demonstrate this new method. The interference
of synchrotron radiation from two undulators was measured over a range
of up to 500\,mm by analyzing the spectrum with a monochromator.  No
coherence loss was observed. The statistical uncertainty could be
reduced to a level of 1\,keV by combining 2\,328 measurements
simultaneously. Systematic uncertainties were dominated by possible
angular misalignments up to 0.15\,mrad. By the use of improved
equipment and alignment techniques, these misalignments should get
controlled on the level of 0.03\,mrad.  The method will then provide
an unprecedented precision of $\delta \gamma/\gamma \le 10^{-4}$
corresponding to an uncertainty in the absolute beam energy $\delta
E_{\text{beam}}$ on the same order of magnitude as the MAMI beam
energy width $\sigma_{\text{beam}}$. The great potential of this
method is achieved by reducing the observables for a beam energy
measurement to a wavelength band in a monochromator and a relative
distance of two light sources in the decimeter range. Such an accurate
energy determination is important for precision hypernuclear physics
experiments on fundamental symmetries. It should also be useful for
precision studies of the accelerator operation and it is worth to note
that other electron accelerators in the energy regime $E_{\text{beam}}
\simeq 200$\,MeV such as S-DANILAC or MESA might benefit from this new
development.

% -------------------------------------------------------------------
\section*{Acknowledgments}
% -------------------------------------------------------------------

We would like to thank the MAMI operators, technical staff, and the
accelerator group for their excellent support of the
experiments. Discussions with M.~Dehn are gratefully acknowledged.

Work supported by Deutsche Forschungsgemeinschaft (DFG) through
Integrated Research Training Group GRK 2128 and Research Grant PO
256/7-1. We acknowledge support by DAAD PPP 57345295/JSPS Research
Cooperative Program and by JSPS KAKENHI No.~JP17H01121 and the
Graduate Program on Physics for the Universe, Tohoku University
(GP-PU).

%--------------------------------------------------------------------- 
%\bibliographystyle{elsarticle-num-names}
%\bibliography{References}
%--------------------------------------------------------------------- 

%=====================================================================
\end{document}